\documentstyle{article}

\def\ben{\begin{equation}}
\def\een{\end{equation}}
\def\bea{\begin{eqnarray}}
\def\eea{\end{eqnarray}}
\input amssym.def
\input amssym.tex
\begin{document}

\hfuzz=100pt
\title{Wormholes on the World Volume:
Born-Infeld particles and Dirichlet p-branes}
\author{G. W. Gibbons
\\
D.A.M.T.P.,
\\ Cambridge University, 
\\ Silver Street,
\\ Cambridge CB3 9EW,
 \\ U.K.}
\maketitle

\begin{abstract}
I describe some recent work in which classical solutions of
Dirac-Born-Infeld theory may be used to throw light on some properties of
M-theory. The sources of Born-Infeld theory are the ends of strings ending
on the world volume. Equivalently the fundamental string may be regarded
as merely a thin and extended piece of the world volume. 
\end{abstract}

\section{Introduction}

In the lecture I described some recent work I had been doing
over a period of time assisted by  my research student 
Dean Rasheed and with some initial assistance from Robert Bartnik.
Some related ideas had been discussed in an 
unpublished note of Douglas,Shwarz and Lowe \cite{DLS}. After the lecture
I was informed by Curt Callan that he and Maldacena had also been thinking
along the same lines. Their work is to be found in \cite{CM} and my own
in \cite{G}. What follows is a slightly extended and expanded version of
what I said in Santiago. The bibliography below is mainly restricted to 
some relevant papers which appeared during the autumn after the lecture.
For a full set of references to earlier work
the
reader is referred to \cite{G,CM,DLS}. 

As is well known, $p$-dimensional extended objects, \lq p-branes\rq
play a central role in establishing the various dualities between the 
the five superstring theories and eleven-dimensional supergravity
theory and these in turn which have  led to the conjecture that there exists a 
single over-arching structure, called \lq M-theory\rq , of which they
may all be considered limiting cases. Whatever M-theory ultimately turns out to be, it is already clear that it is a theory containing p-branes.

Until recently p-branes have been  treated as
\begin{itemize}
\item { Soliton-like BPS solutions of SUGRA theories.}

 or

\item {The ends of open superstrings satisfying
$9-p$ Dirichlet and $p+1$ Neumann boundary conditions.}  
\end{itemize}

My intention is to consider the {\it light brane   approximation}
in which Newton's constant $G\rightarrow 0$ but the string tension
$\alpha ^\prime$ remains finite. In terms of actions we have:

\ben
S= { 1\over g_s^2} S_{NS\otimes NS}^{\rm bulk}  + S_{R\otimes R}^{\rm bulk}  +   
{ 1\over g_s} S^ {\rm brane},
\een
where $S^{\rm bulk}$ is an integral over 10-dimensional spacetime
 and $S^{\rm brane}$ 
is an integral over the $p+1$=dimensional $p+1$ dimensional
world volume of the brane. {\it Heavy branes} correspond to the limit
$g_s \rightarrow \infty$ . {\it Light branes} correspond
to  the limit $g_s\rightarrow 0$.  It is
reasonable to ignore the fields generated by the motion of the brane
and set the Ramond-Ramond fields to zero. 
We therefore consider a Dirichlet p-brane moving in flat
$d+1$ dimensional Minkowski spacetime ${\Bbb E} ^{d,1}$
with constant dilaton and vanishing Kalb-Ramond 3-form.

\section{The Dirac-Born-Infeld action}

For purely bosonic fields  $S^{brane}$ is then given by the 
Dirac-Born-Infeld
action
\ben
-\int _{\Sigma _{p+1}} d^{p+1} x \sqrt{-{\rm det} ( G_{\mu\nu}+F_{\mu \nu} ) }
\een
where 
\ben
G_{\mu \nu}= \partial_\mu  Z^A \partial _\nu Z^B \eta_{AB}
\een
is the pull back of the Minkowski metric $\eta_{AB}$ to the world volume
$\Sigma _{p+1}$ using the embedding map $Z^A(x^\mu): \Sigma_{p+1} \rightarrow {\Bbb E} ^{d,1}$ and
\ben
F_{\mu \nu}=\partial _\mu A_\nu - \partial_\nu A_\mu
\een
is the curvature or field strength of an abelian connection
$A_\mu(x^\nu)$ defined over the world volume. 

The Dirac-Born-Infeld action is invariant under the semi-direct product
of  
\begin{itemize}
 \item{ world volume diffeomorphisms}
 and
\item{ abelian gauge transformations.}
\end{itemize}
To fix the former we adopt {\it Static Gauge}, called by mathematicians the {\it non-parametric representation}
:
\ben
Z^M=x^\mu, M=0,1\dots, p
\een
\ben
Z^M=y^m, M=p+1, \dots,d-p.
\een
The transverse coordinates $y^m$ behave like scalar 
fields on the world volume and the action becomes
\ben
-\int_{\Sigma _{p+1}}  d^{p+1} x \sqrt {-{\rm det}( \eta_{\mu\nu} +\partial_ \mu y^m \partial _\nu y^m+F_{\mu \nu}) }.
\een
Note that  original manifest global Poincar\'e symmetry $E(d,1)$ has been
reduced to a manifest $E(p,1) \times SO(d-p)$.

An important message of this work is that static gauge cannot usually
be globally well defined and it generates spurious singularities
if the world volume is topologically non-trivial. Even if $\Sigma_{p+1}$
is topologically trivial, static gauge may still break down if the
brane bends bcak on itself. Geomtrically we have projected $\Sigma _{p+1}$
onto a $p+1$ hyperplane in ${\Bbb E}^{d,1}$ 
and the $y^m$ are the height functions. However the projection need
not be one-one.

Static gauge makes apparent that there are two useful 
consistent truncations
\begin{itemize}
 
\item{ $y^m=0$ which is pure Born-Infeld theory\cite{BI,B} in ${\Bbb E}^{p,1}$} 

 and
 
\item{ $F_{\mu \nu}=0$ which corresponds to Dirac's
theory\cite{D}  of minimal timelike
submanifolds of ${\Bbb E}^{d,1}$.}
\end{itemize}
The basic time independent solutions in these two cases are the BIon
and the catenoid respectively. We shall see that there are two 
one-parameter families of solutions interpolating between them,
rather analogous to the one parameter family of Reissner- Nordrstrom
black holes. An internal Harrison like $SO(1,1)$ boost
symmetry
moves us along the two families, one of which is \lq sub-extreme\rq
and the other of which is \lq super-extreme\rq.
The two families are separated by an extreme BPS type solution. 

\section {BIons}

Let's start with Born-Infeld theory and consider the case
$p=3$. Similar results hold if $p\ne 3$. 
The original aim of this theory was to construct classical finite energy
pointlike solutions representing charged particles. It
is these that I  call \lq BIons\rq and their study \lq BIonics\rq.

For time independent pure electric solutions
the lagrangian reduces to
\ben
L=-\sqrt{ 1-{\bf E}^2}+1.
\een
where ${\bf E}= -\nabla \phi$ is the electric field.
Thus the electric induction is 
\ben {\bf D}= {\partial L \over \partial {\bf E}}=
{{\bf E} \over \sqrt{1-{\bf E}^2} }
\een
and thus
\ben
{\bf E}= {{\bf D} \over \sqrt{ 1+ {\bf D}^2 } }.
\een
Now if 
\ben
\nabla \cdot {\bf D} = 4 \pi e \delta ({\bf x}),
\een
\ben
{\bf E}= { e {\hat {\bf r}} \over \sqrt{ e^2 + r^4} }.
\een
Clearly while the induction ${\bf D}$ diverges at the origin
the electric field remains bounded and attains unit magnitude
at the origin. In other words the slope of the potential is 
45 degrees at the origin.

The energy density is $T_{00}= {\bf E} \cdot {\bf D}-L$
and it is easy to see that the total energy is finite.

It is important to realize the difference ,
not widely understood, between  \lq BIons\rq and
conventional \lq solitons\rq. Originally Born-Infeld theory was intended 
as a \lq unitary\rq theory of electromagnetism. In modern terms 
such a theory would be one in which the classical electron
is represented by an everywhere non-singular finite energy of 
the source free non-linear equations of motion.  
In such theories the particle equations of motion follow from
the equations of motion of the fields without having to be postulated
separately. As such the theory was a failure 
because
\begin{itemize} \item{ The BIon solutions have {\sl sources }.}
 \item{The solutions are still {\sl singular} at the location of the source.}

and

\item{ One must impose {\sl boundary conditions} on the singularities in order to obtain the equations of motion. For a recent 
discussion of this point see\cite{C,CC}.}
\end{itemize}
Nevertheless \lq BIonic\rq solutions of field theories frequently have a 
sensible physical interpretation (cf. point defects in liquid crystals).
To illustrate the point we consider briefly Born-Infeld electrostatics.
Solutions of the  equation of motion
\ben
\nabla \cdot { \nabla \phi \over \sqrt{ 1- |\nabla \phi|^2 } }
\een
may be interpreted as spacelike maximal hypersurfaces $t=\phi({\bf x})$
in an auxiliary
$p+1$ Minkowski spacetime with coordinates $(t,{\bf x})$. This allows
one to use geometrical techniques from general relativity and p.d.e. theory
to discuss the existence and uniqueness of solutions.
More significantly it allows us to construct new solutions.
For example boosting the trivial solution with velocity $v=E$ 
in the $z$ direction gives rise
to a uniform electric field $\phi=-Ez$. The maximal field strength
in Born-Infeld theory corresponds to the maximum velocity in special
relativity. 

One may also boost the BIon solution to give a {\sl static} solution
representing a charged particle at rest in an asymptotically uniform
electrostatic field $E$. This sounds paradoxical but it is not.
The point is that the solution does not satisfy the correct
boundary conditions at the particle centre to be force free. 
It is pinned by a force $F=eE$ given by 
\ben
F_i= \int T_{ij} d\sigma _j
\een
where the integral is taken over a sphere surrounding the Bion.

Static solutions which {\sl do} satisfy the force free solution
can be also found. Thus if ${\frak{p}(x)} $ 
is the Weirstrass elliptic function with
$g_3=0$ and $g_2=4$ then
\ben
{\frak{p}(x)} {\frak{p}(y)} {\frak{p}(z)} ={\frak{p}(\phi)} 
\een
we get a BIon crystal of NaCl type. In this case the forces on the BIons
cancel by symmetry. In general one may apply comparative statics
and the virial theorem to obtain some striking analogues of the
results in black hole theory. If ${\bf F}^a$ is the force on the $a$'th
BIon which has position ${\bf x}_a$, charge $e_a$ and electrostatic 
potential $\Phi^a$ one has the \lq second law\rq 
\ben
dM=\Phi ^a de_a + {\bf F} ^a \cdot  d {\bf x}_a
\een
and the Smarr-Virial relation:
\ben
M={ 1\over 3} {\bf F} ^a  \cdot {\bf x}_a + { 2\over 3} e_a \phi ^a.
\een
Here $M$ is the total energy  and there is a sum over the BIon
index $a$.

\section {Catenoids}

Consider one transverse coordinate $y$. The lagrangian now becomes
\ben
L= -\sqrt{ 1+ |\nabla y|^2}+1.
\een
If $p=3$ We soon find that a spherically symmetric solution satisfies
\ben
\partial _ry= \pm { c \over \sqrt {r^4-c^2} }
\een.

The solution breaks down at $r=\sqrt{c} $ because of a breakdown
of static gauge. In fact the spatial part of the world volume
(i.e. the p-brane) consists of two copies of the solution
for $r>\sqrt{c}$ joined by a minimal throat. In other words, the 
solution has the geometry of the  Einstein-Rosen throats
familiar in Black Hole theory. (In fact the Einstein Rosen throats,
i.e. the constant time surfaces of static black holes 
or of self-gravitating p-branes, {\sl are} 
minimal submanifolds). 

Near infinity the catenoid looks like two parallel $p$-planes
situated a finite distance $Y$ apart. Callan and Maldacena\cite{M}
have suggested that one should regard this as a D-brane-anti-D-brane
configuration though it is a single connected surface.
The catenoid is unstable in that it one can find a deformation which lowers
the total volume. For that reason it was suggested by them that it should be
thought of as some sort of sphaleron. 
It is interesting to note that subsequent 
to, and independent of, their discussion
there appeared a paper \cite{CSZ} in which
the it was shown, using the fact that
 that the Hessian ( i.e second variation
) of the
Dirac energy is
\ben
\int \sqrt{g} d^{p} x f \Bigl( -\nabla ^2 _g -K_{ij} K^{ij} \Bigl ) f.
\een
where $K_{ij}$ is the second fundamental form of the
hypersurface,   
that  quite generally any complete minimal hypersurface  of 
${\Bbb E}^{p+1} $ with more than one end admits
bounded harmonic functions and thus cannot be a true minimum
of the energy.

\section{Charged Catenoids}

In general the relevant lagrangian is

\ben
L=-\sqrt{ 1-|\nabla \phi|^2 +|\nabla y|^2 + (\nabla y \cdot \nabla \phi)^2- (\nabla \phi )^2 ( \nabla y)^2 }
\een
This is manifestly invariant under  generalized Harrison
transformations consisting of boosts in the $\phi-y$ plane.
Starting from the Bion or the catenoid we obtain the two one parameter
families mentioned above. Note that the super-extreme solutions have a 
singular source on the world volume 
while the sub-extreme solutions are perfectly regular and have no
source on the world volume. Starting with the catenoid and charging
it up gives a narrower and narrower and longer and longer throat.
Starting with the BIon and adding the scalar gives a bigger and bigger spike.
The interesting question is what about the limiting case?

\section{The BPS solution}

It is a simple task to verify that 
taking
\ben
\phi=\pm y =H
\een
where $H$ is an arbitrary harmonic function will solve the equations.
If
\ben
H= \sum _a { c_a \over |{\bf x}- {\bf x}_a | },
\een
we  get a superposition of arbitrarily many infintely spiky solutions.
One may  verify that these solutions are supersymmetric
and indeed they satisfy the effective equations of motion of the
superstring to all orders \cite{T}. We shall see why shortly.
In the mean time we point out that the obvious
natural interpretation of these
solutions is that they represent infinitely long
fundamental strings ending on
a D-brane as first envisaged by Strominger and by Townsend.
We can now see clearly where the source for the BIon comes from. It is
carried by the string. Indeed one may check that the
charge carried by the string equals that carried by the BIon 
using the fact the the coupling to the Neveu-Scharz field $B_{\mu \nu}$
in the Dirac-Born-Infeld action is obtained by the replacement
\ben
F_{\mu \nu} \rightarrow {\cal F} _{\mu \nu} = F_{\mu \nu} - B_{\mu \nu}.
\een

Note that if $c=1$ the D-brane spike has height $L$ at a distance 
$ 1\over L$
from the source.
The paper by Callan and Maldacena\cite{CM}, see also \cite{H}
 gives more detailed evidence
for this viewpoint by showing that,  by being careful about
factors, the energy of a length  $L$ of  string
agrees with the world volume energy of the fields 
outside a radius $1 \over L$.

\section {Electric-magnetic duality and the inclusion of magnetic fields}

We have  ${\bf H}= - {\partial L \over \partial {\bf B}}=-\nabla \chi$
where $\chi$ is the magnetostatic potential. 
Let $\Phi^A = (y,\phi,\chi)$ be coordinates in an 
auxilliary Minkowski spacetime ${\Bbb E}^{1,2}$ ( with two negative signs).

By means of a suitable Legendre transformation one may obtain an 
effective action from which to deduce the equations of motion. It is
\ben
\sqrt{ {\rm det} ( \nabla \Phi^ A \cdot \nabla \Phi^B - \eta ^{AB} ) }.
\een

This is manifestly invariant under $SO(2,1) \supset SO(2)$.
The $SO(2)$ subgroup of rotations of $\phi$ into $\chi$ is of course
just electric-magnetic duality rotations. It is well known that
Born-Infeld theory has this symmetry. In fact its existence
 may be traced back to
the basic S-duality of non-perturbative string theory.
Acting on the solutions with it we can obtain magnetically charged BIons
attached to D-strings. In fact classically there is an entire circle
of dyonic BPS solutions but of course 
quantization breaks down $SL(2,{\Bbb R})$ to $SL(2,{\Bbb Z})$.

\section{ Abelian Bogomol'nyi Monopoles}

Setting $\phi=0$ we find magnetic solutions with ${\bf B}={\bf H}$.
Thinking of $y$ as a Higgs field we recognize the equations
\ben
\nabla y= \pm {\bf B}
\een
as the abelian Bogomol'nyi equations of Yang-Mills theory, valid in the limit that
the mass $m_W$ of the  vector bosons goes to infinity.
This is consistent with Witten's ideas about nearby Dirichlet-branes.
If two branes are well separated one has  a gauge group $U(1) \times U(1)$,
one of the factors corresponding to the centre of mass motion. 
If they are coincident one expects symmetry enhancement to $U(2)$.
Taking out a $U(1)$ factor corresponding to the centre of mass 
the world volume gauge group is $SU(2)$. The distance $Y$ between
the branes is supposed to be proportional to $m_W$. 
The abelian Born-Infeld theory does indeed
seem able to capture the physics of the large vector-boson mass limit.

\section{Dimensional Reduction and SUSY}

It is rather convenient to obtain the Dirac-Born-Infeld lagrangian
in static gauge by dimensionally reducing the pure Born-Infeld lagrangian
\ben
-\sqrt{ -{\rm det} (\eta_{AB} + F_{AB})}
\een
from ten dimensions to $p+1$ dimensions. One sets
\ben
A_A= (A_m(x), A_\mu(x))
\een
and identifies the transverse components $A_m$ of 
the gauge connection one-form $A_A$ with the transverse coordinates $y^m$ of the $p+1$ brane.
At lowest order \, one may use the supersymmetry transformations
of 1-dimensional SUSY (abelian) Yang-Mills. Thus SUSY requires the existence of a sixteen component Majorana-Weyl Killing spinor
$\epsilon$ such that
\ben
F_{AB} \gamma ^A \gamma ^B \epsilon=0.
\een
In the electric case, our ansatz is
\ben
F_{5i}=F_{0i},
\een
so the BPS condition requires $(\gamma ^0+\gamma^5)\epsilon=0$, which has
eight real solutions. The self dual solutions are also easily seen
to be BPS. One may check that both continue to admit killing spinors
when the full non-linear supersymmetry transformations have been 
taken into account \cite{HLW}. Moreover, in the electric case
it has been argued  that the 
solution gives an exact boundary conformal field theory\cite{T}.
This is almost obvious  because of the lightlike nature
of the the electric ansatz. All contractions involving $F_{AB}$
must vanish.

The easily verified fact that the abelian
anti-self-duality equations are sufficient conditions  
for solutions the Born-Infeld equations leads to an interesting and useful
relation to minimal 2-surfaces in ${\Bbb E}^4$. One assumes that $A_\mu$ 
depends only on $z=x^1+ix^2$. Setting $A_3+iA_4=w=x_3+ix_4$ one
then easily calculates that $F_{\mu \nu}=-\star F_{\mu \nu}$ reduces 
to the Cauchy-Riemann equations, i.e. 
to the condition that $w$ is a locally holomorphic function of $z$.
In this way one may obtain a variety of interesting minimal 
surfaces which can in fact be regarded as exact solutions of M-theory.
Thus if 
\ben
wz=c,
\een
and $c\ne0$, we obtain a smooth connected
2-brane with topology 
${\Bbb C}^\star \equiv {\Bbb C} \setminus 0$ looking like two
2-planes connected by a throat. If $c=0$ this degenerates
to two 2-planes,
$z=0$ and $w=0$ intersecting at a point.
Constructions of minimal surfaces in ${\Bbb E}^4$ using holomorphic
embeddings were pioneered by Kommerell in 1911 \cite{K}
so it seems reasonable
to refer to them as Kommerell solutions. The reader is referred \cite{M} for
an account of  recent applications to gauge theory.

\section {Calibrated geometries}

The holomorphic solutions of the last section
are in fact a special case
of a more general class of solutions in which the familiar Wirtinger's
inequality for the area, or more generally
the p-volume, of holomorphically embedded p-cycles
is replaced by a more general inequality. The basic idea is to replace
a suitable power of the K\"ahler form by some other closed p-form.
The form is called by Harvey and Lawson \cite{HL} a calibrating p-form
Using their work one see that  the Dirac-Born-Infeld  equations have a very rich
set of solutions. Here is not the place to discuss them
and their applications to gauge theory
in detail.
I will simply remark that one encounters
various kinds of topological defects
on the world volume, such as vortices and global monopoles. These latter
may be relevant to discussions of the non-abelian
monopoles in the limit that the gauge field decouples.
Of special interest are the BPS solutions and their Bogomol'nyi
bounds. It turns out that there is a close connection between this,
kappa symmetry,  and the calibration condition 
 of Harvey and Lawson\cite{HL}. In fact he calibration condition turns 
out to be the condition for supersymmetry\cite{GP}.

\section{Non-Abelian Born-Infeld}

It is widely believed that when a number of $D$-branes
coincide there is symmetry enhancement. The current
most popular suggestion for the relevant generalization
of the Born-Infeld action is that of Tseytlin
\ben
- {\rm Str} \sqrt{ -{\rm det} (\eta_{AB} + F_{AB})}
\een
where $F_{AB}$ is in the adjoint representation of the gauge group $G$
and $\rm Str$ denotes the symmetrized trace of any product of
matrices in the adjoint representation that it precedes.
The Tseytlin action has the property,
\cite{GT,H} that solutions of the non-abelian Bogomol'nyi
equations are also solutions. In the case of $SU(2)$ one may even establish
a generalized Bogomol'nyi bound \cite{GT}. Very recently\cite{GGT}
BPS bounds have also been established in 
a rather different way using an apparently  different energy functional.

\section{Conclusion} 

A  striking aspect of the work reported above is the way in which
Born-Infeld theory has finally
found a home on the world-volume
and its mysterious sources 
have been shown 
to be just the ends of strings extending into higher dimensions.
Even more striking is the way in which  
the fundamental string solution emerges as a limiting case of
M-theory solutions. It seems to 
reinforce the widely held viewpoint that in the ultimate
formulation of the theory, strings as such may have no fundamental
role to play and may indeed appear only as effective excitations. 
However, as always,
it is worth exercising some caution. After all, who would have thought
a few years ago that Born-Infeld theory 
and Dirac's doomed attempt to construct an extended model of the electron,   
long since relegated to the dustbin of history and condemned
as a  last nostalgic gasp
at the fag-end of the classical world-picture should 
re-emerge at the cutting edge
of post modernist physics?

There will no doubt be many more 
fag-ends and even a few cigars before the final story is told. In the meantime
it is my pleasant duty to thank Claudio and Jorge, so ably assisted by the Chilean
Air Force, 
for organizing such a wonderful conference and making our stay in Chile and Antartica so memorable.

\end{document}